\documentclass[epj]{svjour}
\usepackage{graphics}
\usepackage{amsfonts}
\usepackage{amsbsy}
\usepackage{mathrsfs}
\usepackage{amsmath}
\usepackage{amssymb}
\begin{document}
\title{On parton distributions in a photon gas}
\author{I. Alikhanov\thanks{\email{ialspbu@gmail.com}}
}                     
\institute{Institute for Nuclear Research of the Russian Academy
of Sciences, 60-th October Anniversary pr. 7a, Moscow 117312, Russia}
%
%
\abstract{In some cases it may be useful to know parton distributions in a photon gas.
This may be relevant, e.g., for the analysis of interactions of high energy
cosmic ray particles with the cosmic microwave background radiation. The latter
can be considered as a gas of photons with an almost perfect blackbody
spectrum. An approach to finding such parton distributions is described. The survival probability of ultra-high energy neutrinos traveling through this radiation is calculated.
} 
\authorrunning{I. Alikhanov}
\titlerunning{On parton distributions in a photon gas} \maketitle
%
%
\section{Introduction}{\label{introd}}
There are reactions in which photons manifest hadronic
properties~\cite{feynman}. In an analysis of such reactions it may be
possible to employ formalism used for the case of hadrons, at least
partly. For example, one can introduce the structure function
characterizing the parton densities in the
photon~\cite{review_phot1,review_phot2}. It was suggested in the early 1970's that information on this
structure function might be accessible by deep
inelastic electron--photon scattering at $e^+e^-$ colliders~\cite{ref:dis_photon1,ref:dis_photon2}.
Theoretical investigations at that time already revealed the basic logarithmic dependence of
the function on the four-momentum transfer squared $Q^2$~\cite{ref:log_dep1,ref:log_dep2}. 
Quantum-chromodynamics (QCD) corrections
to the pointlike structure of the photon were
calculated~\cite{ref:qcd}.  Evolution equations for the parton
densities as well as the properties of the corresponding solutions
were under scrutiny (see, e.g.,~\cite{gluk3}). A description of
recent developments and a more complete list of references can be
found, for example,
in~\cite{review_phot1,review_phot2,review_phot3,review_phot4}.

The universe is filled with the cosmic microwave
background (CMB) radiation which can be considered as a gas of photons
with an almost perfect blackbody spectrum at temperature $T\approx2.725$ K~\cite{cmb_spectr}.
The CMB plays an important role in astrophysics, for
example, providing a medium which inevitably interacts with high energy cosmic ray
particles~\cite{greisen,zatsepin,sigl}.
We have recently pointed out that owing to the existence
of this radiation the parton content of the real photon
may also find a non-trivial astrophysical application ~\cite{cmb_parton}.
It may be therefore useful to know parton distributions associated with the CMB.

The typical energy of a photon of the CMB at the temperature 2.725 K
is very low, about $10^{-3}$ eV. Nevertheless one should not be
embarrassed by such energies. In fact, not every reference frame is
suitable for an approach based on the parton model. There are frames
in which the parton densities are undefined, as for example in the
rest frame of a particle whose partonic structure is studied. One
can, however, successfully use the parton model in a coordinate
system with respect to which the particle is moving with very high
momentum (this question is nicely discussed, e.g., in
\cite{greiner}). Thus, a reference frame fixed to a high energy
cosmic ray particle propagating through the universe and interacting
with the CMB may be an appropriate one.

%
\section{Derivation of parton distribution functions}{\label{section2}}
Let us consider a photon gas with blackbody spectrum. The number of
photons in volume $V$ at temperature~$T$ with energies between
$\omega$ and $\omega+d\omega$ is given by~\cite{landau}

\begin{equation}
N(\omega,
T)\mathrm{d}\omega=\frac{V}{\pi^2(c\hbar)^3}\,\frac{\omega^2\mathrm{d}\omega}{e^{\omega\hskip
-0.5mm/\hskip -0.4mm kT}-1},\label{eq:planck}
\end{equation}

where $k$ is Boltzmann's constant, $c$ is the speed of light and
$\hbar$ is the reduced Planck constant. Henceforth we set
$c=\hbar=1$. Note that (\ref{eq:planck}) is a Lorentz invariant.

Dividing  (\ref{eq:planck}) by the total number of photons

\begin{equation}
N=\frac{2\zeta(3)}{\pi^2}\,Vk^3T^3 \label{ntot}
\end{equation}

(where $\zeta(s)$ is the Riemann zeta-function \cite{zeta}) yields
the volume-independent probability to find a photon in this energy
interval

\begin{equation}
n(\omega,
T)\mathrm{d}\omega=\frac{1}{2\zeta(3)k^3T^3}\,\frac{\omega^2\mathrm{d}\omega}{e^{\omega\hskip
-0.5mm/\hskip -0.4mm kT}-1}. \label{eq:norm}
\end{equation}

One can see that (\ref{eq:norm}) is only a function of the
ratio

\begin{equation}
x=\frac{\omega}{kT} \label{eq:var_x}
\end{equation}

and may be rewritten as

\begin{equation}
n(x)\mathrm{d}x=\frac{1}{2\zeta(3)}\,\frac{x^2\mathrm{d}x}{e^{x}-1}.\label{eq:planck_norm}
\end{equation}

In order to make the subsequent discussion more customary, let us
regard, just formally, $n(x)$ as the probability density function to
find a photon carrying the "fraction"~$x$ of the energy $kT$. We
emphasize that $x$ may vary,  in the range from zero to infinity (this
is the reason of using the quotation marks).

We define the sought-for parton distributions as functions of this
dimensionless variable $x$ in the following way:

\begin{equation}
q(x,Q^2)= \int_x^\infty\frac{\mathrm{d}y}{y}\,n(y)\, {q}^{\gamma}\hskip
-1mm\left(\frac{x}{y},Q^2\right); \label{eq:final_result}
\end{equation}

here ${q}^{\gamma}(\xi,Q^2)$ parametrizes the probability density to
find a quark of flavor $q$ with the momentum fraction $\xi$ in the
photon probed by a hard scattering at virtuality scale~$Q^2$.

Though $x$ and $y$ in general vary from zero to infinity, their
ratio $x/y$ in (\ref{eq:final_result}) is always less than or equal
to unity and the function ${q}^{\gamma}(\xi,Q^2)$ retains its usual
meaning. We adopted the result of the quark--parton model
\cite{review_phot1}:

\begin{multline}
q^{\gamma}(\xi,Q^2)=N_c\hskip 0.15mm e^2_q\frac{\alpha}{2\pi}\Biggl[8\xi(1-\xi)-1\Biggr.\\+
\Biggl.[\xi^2+(1-\xi)^2]\ln\left(\frac{Q^2}{m_q^2}\frac{1-\xi}{\xi}\right)\Biggr],
\label{eq:qpm}
\end{multline}

where $N_c$ is the number of the quark colours, $\alpha$ is the fine structure constant, $e_q$ and $m_q$ are
the electric charge and mass of the quark $q$, respectively.

\begin{figure}
\centering
\resizebox{0.5\textwidth}{!}{%
\includegraphics{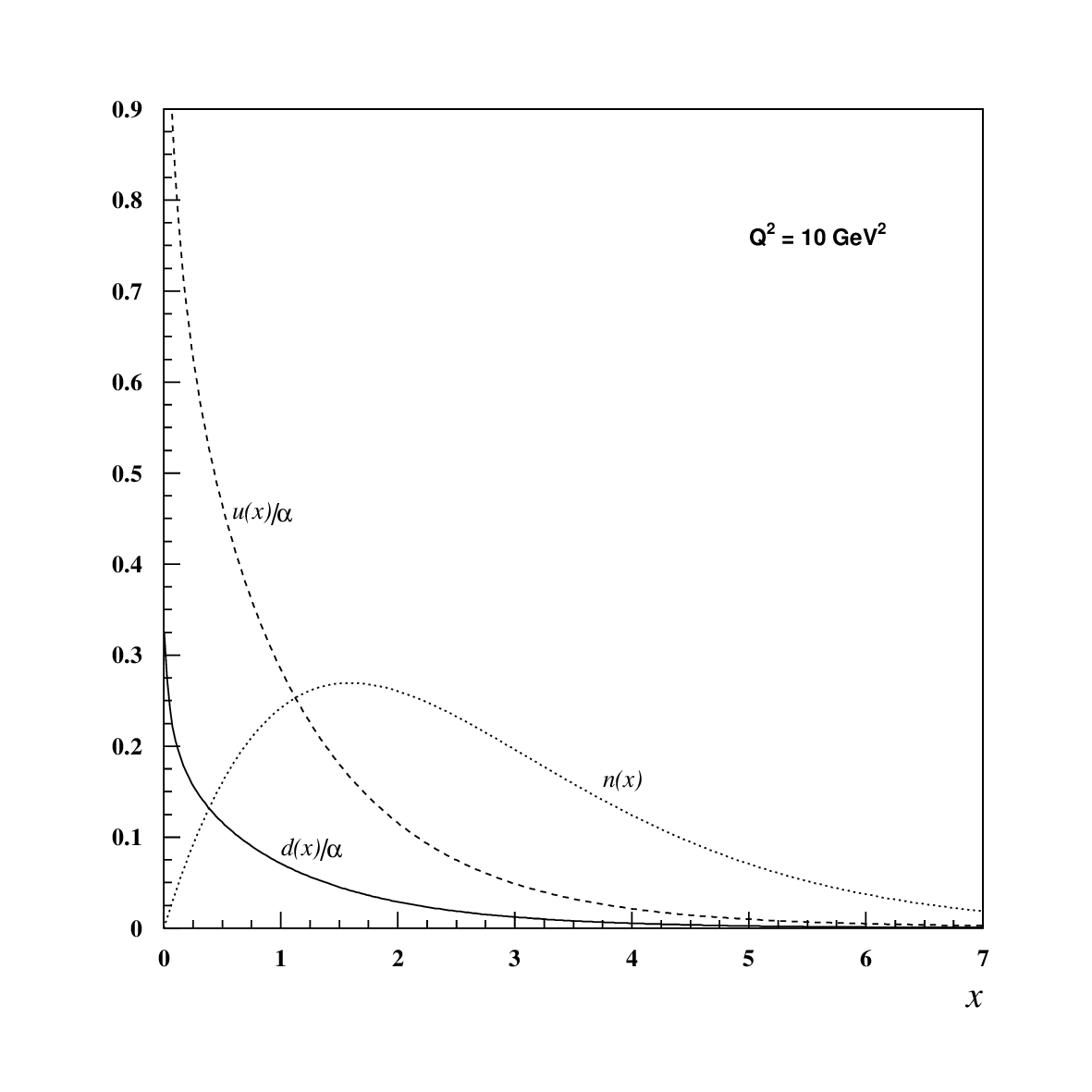}
}\caption{The parton distributions in the photon gas as
functions of~$x$: $u$~quark (dashed curve) and $d$~quark (solid
curve). Note that $Q^2$ is fixed at 10~GeV$^2$. The distribution of
the photons $n(x)$ is also displayed by the dotted curve}
\label{fig1}
\end{figure}

Using (\ref{eq:planck_norm}) and (\ref{eq:qpm}), we have numerically
solved (\ref{eq:final_result}) for $u$ and $d$ quarks. We set $N_c=3$,
$Q^2=10$ GeV$^2$ and $m_u=m_d=0.2$~GeV. The obtained results are
depicted in Fig.~\ref{fig1}. To make the picture more complete, the
distribution of the photons $n(x)$ is also displayed. One can see
that it is more probable to find the quarks with small values of
$x$, while the number of photons decreases as $x$ approaches zero.
The corresponding antiquarks behave analogously.

In principle, more precise calculations may be performed when
needed, for example, by using more realistic parton densities
instead of (\ref{eq:qpm}). The possibility of a presence of strange
and heavier quarks in the photon can additionally be taken into
account. But the main task of this paper is to give a qualitative
description of the approach itself. For this reason we did not show
the $Q^2$-dependence of the parton distributions in detail (this is
weak, being logarithmic, and does not qualitatively change the
results).

We note the Lorentz invariance of this formalism which follows from the fact that (\ref{eq:planck})
and (\ref{eq:qpm}) are  defined in a Lorentz-invariant way.

One may also derive distributions of the charged leptons  $f_{l}(x,Q^2)$ exactly as above, just by making the following changes in (\ref{eq:qpm}): $N_c\to1$, $e_q\to1$, $m_q\to m_l$ \cite{review_phot1}, where the subscript $l$ refers to the leptons.

\begin{figure}
\centering
\resizebox{0.5\textwidth}{!}{%
\includegraphics{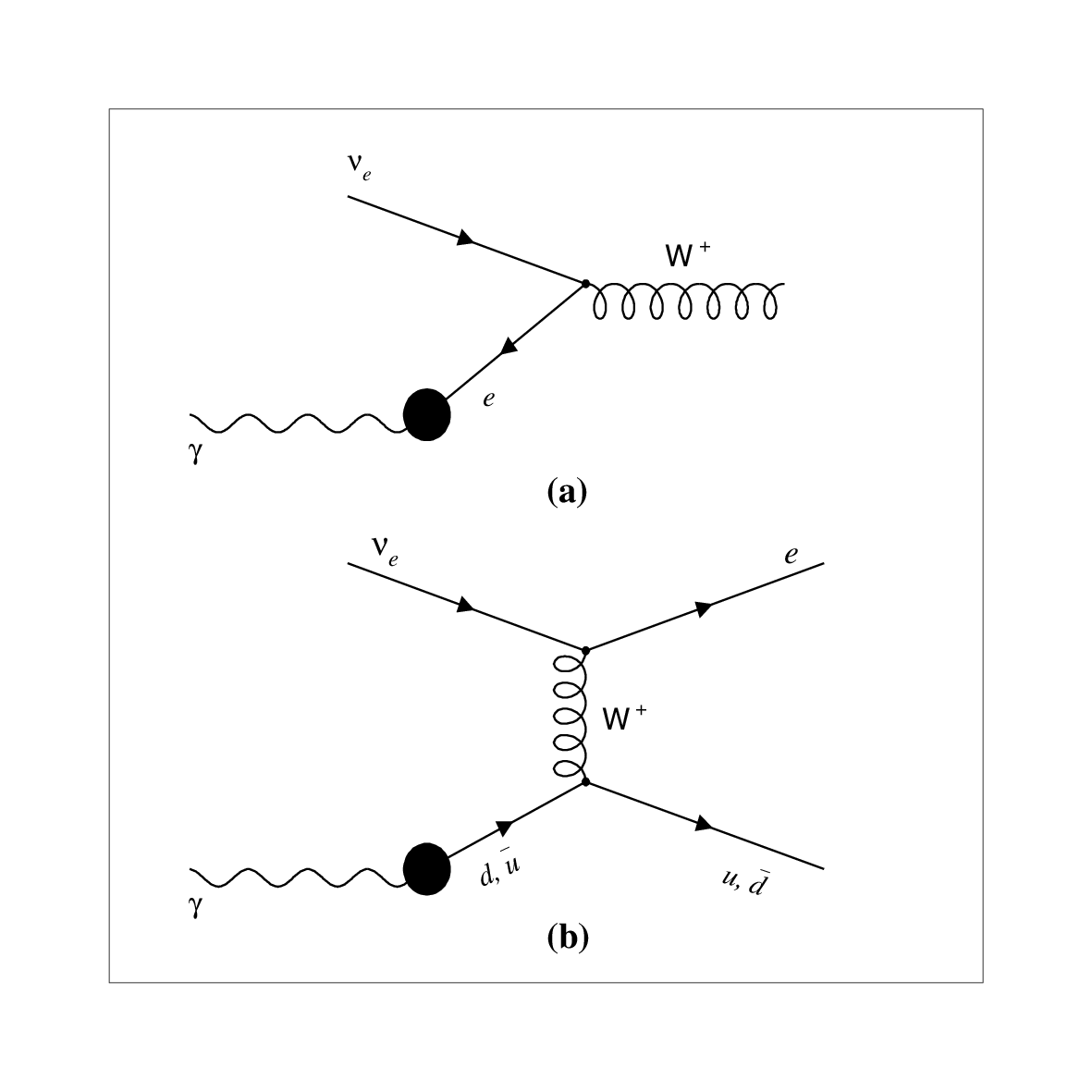}
}\caption{Diagrams illustrating {\bf a} the inclusive on-shell $W^+$ boson production $\nu_e\gamma\rightarrow W^+X$;
{\bf b} charged current neutrino scattering off quarks (antiquarks). In this paper we take only the $u$ and $d$ quarks (antiquarks) into account and neglect Cabibbo--Kobayashi--Maskawa mixing}
\label{fig2}
\end{figure}

\section{Neutrino absorption by the CMB radiation}{\label{section3}}
Here we illustrate how the model presented in the previous section can be applied.
For this purpose we consider two possible channels of absorption of ultra-high energy (UHE) neutrinos by the CMB radiation, namely the inclusive resonant $W^+$ boson production

\begin{equation}
\nu_e\gamma\rightarrow W^+X\label{s_channel}
\end{equation}

and the neutrino--photon deep inelastic scattering

\begin{equation}
\nu_e\gamma\rightarrow e^-X.\label{t_channel}
\end{equation}

The corresponding diagrams are shown in Fig. \ref{fig2}. A~channel closely related to (\ref{s_channel}) has been proposed in~\cite{seckel}.
Let us discuss only the case of the electron neutrino since the other flavours can be treated likewise.  

It is obvious that the incident neutrino has to possess extremely high energy to interact with the CMB photons via these channels \cite{cmb_parton}. This is similar to the situation when one needs to probe the region of small values of the Bjorken variable $x_{_B}$ in the nucleon. As objects of QCD its valence quarks are surrounded by a cloud of soft virtual gluons which may generate quark--antiquark pairs dominant at $x_{_B}\ll1$. The lower values of $x_{_B}$ are to be studied, the higher energies in the center-of-mass system of the colliding nucleons are required. Analogously we regard the CMB radiation as such a cloud whose quark component can be resolved by the UHE neutrinos.

\begin{figure}
\centering
\resizebox{0.5\textwidth}{!}{%
\includegraphics{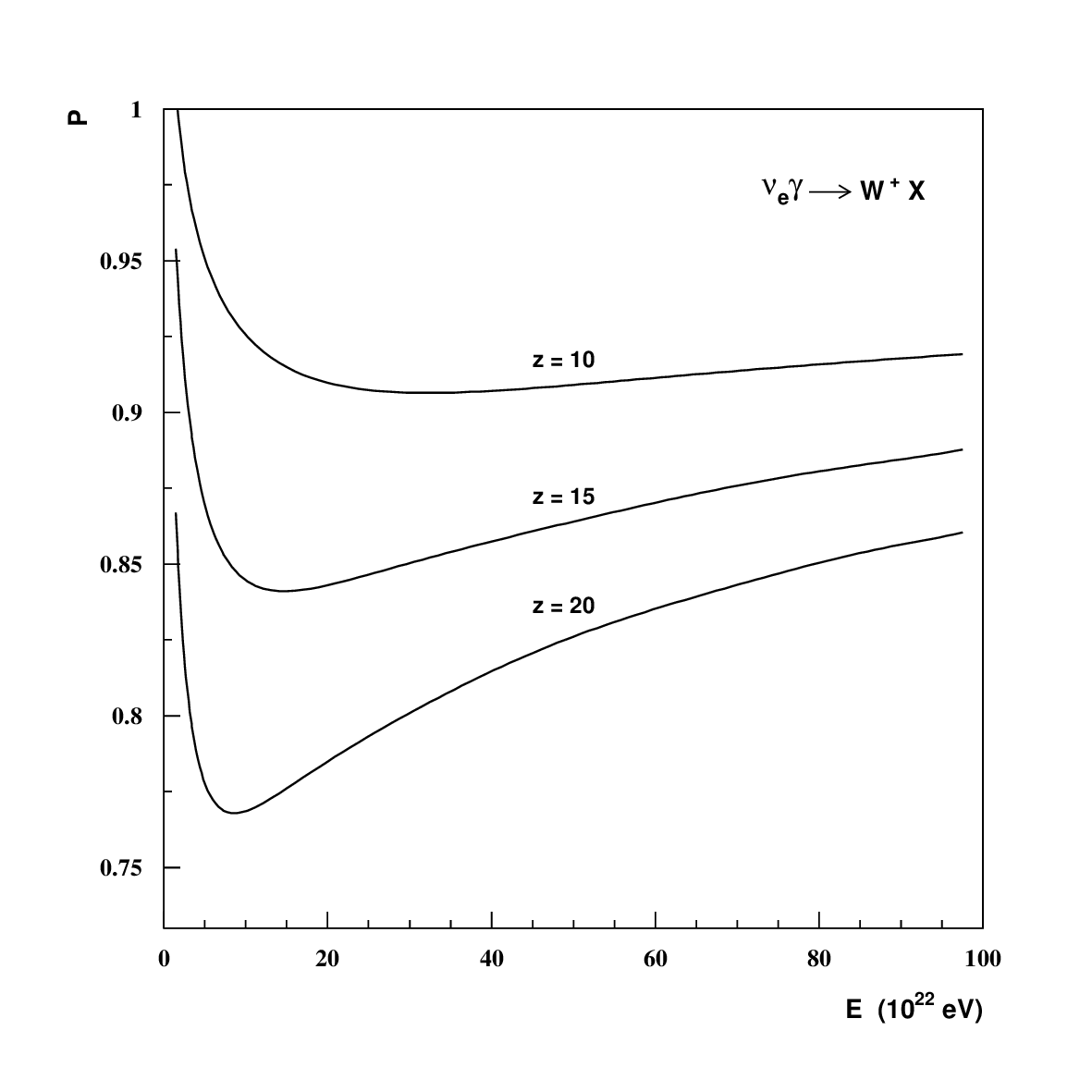}
}\caption{Dependence of the survival probability of an UHE electron neutrino emitted at $z_s=10, 15$ and $20$ on its present day  energy $E$. The absorption channel is $\nu_e\gamma\rightarrow W^+X$}
\label{fig3}
\end{figure}

There have been works devoted to a similar problem -- the problem of damping of the UHE neutrinos in the cosmic relic neutrino background, for example \cite{neutr_damping1,neutr_damping2,neutr_damping3,neutr_damping4}. Following them we will calculate the damping rate of the neutrinos in the CMB.

It is convenient to begin with a consideration of the $\nu\gamma$ interactions at the subprocess level explicitly writing down some kinematical quantities. Thus, let $\sigma(\hat s, Q^2)$ be the total cross section of a neutrino--parton scattering, where standardly

\begin{equation}
\hat s=2E\varepsilon(1-\cos\theta)\label{cms_energy};
\end{equation}

here $E$ and $\varepsilon$ are the energies of the incident neutrino and parton, respectively (their masses are neglected), $\theta$ is the angle between the momenta of the colliding particles. Note that $Q^2$ is the same at the neutrino--photon and neutrino--parton levels \cite{close}.

For a static target the damping rate $\gamma$ would simply read $\gamma=\sigma n$ (recall that $c=1$) with $n$ and $\sigma$ being the number of target particles per unit volume and cross section of a given reaction, respectively. In our case, however, the scatterers are not at rest and we have to take their energy distribution into account. We define the damping rate at fixed $Q^2$ and $T$ in the form

\begin{equation}
\gamma(E)=\int \mathrm{d}\varepsilon\int\limits_{\omega\geq\varepsilon}\frac{\mathrm{d}^3\vec\omega}{{(2\pi)}^3\omega}\frac{2q^{\gamma}(\varepsilon/\omega,Q^2)}{e^{\omega\hskip
-0.4mm/\hskip -0.4mm kT}-1}\sigma(\hat s,Q^2), \label{eq:damping_def}
\end{equation}

where $\sigma(\hat s,Q^2)$ is included in the integral over $\omega$ since it depends on the angle $\theta$ that also enters into the momentum space element $\mathrm{d}^3\vec\omega$ (in other words, the parton originates from the photon and the momenta of these particles are assumed to point in the same direction). The integration over $\varepsilon$ is performed over kinematically allowed parton energies. One can see that in (\ref{eq:damping_def}), additionally to the familiar Planck's distribution $2/[e^{\omega\hskip
-0.4mm/\hskip -0.4mm kT}-1]$, we have introduced the function $q^{\gamma}(\varepsilon/\omega,Q^2)/\omega$ to describe photon splitting into quark--antiquark pairs in a fashion analogous to the parton model. Actually, using that $\mathrm{d}^3\vec\omega=2\pi\omega^2\mathrm{d}\omega\sin{\theta}\mathrm{d}\theta$ and making the change of variables according to (\ref{eq:var_x}) ($x=\varepsilon/kT$, $y=\omega/kT$) in  (\ref{eq:damping_def}) lead to

\begin{multline}
\gamma(E)=\frac{n_0}{2}\int \mathrm{d}x\left\langle\sigma(xs,Q^2)\right\rangle\\\times\int_x^\infty\frac{\mathrm{d}y}{y}\,n(y)\, {q}^{\gamma}\hskip
-1mm\left(\frac{x}{y},Q^2\right), \label{eq:final_result1}
\end{multline}    

where the brackets $\langle\ldots\rangle$ indicate the standard integration over the polar angle $\theta$, $s=2EkT(1-\cos\theta)$, $n_0=N/V$ is the number density of the CMB photons (see (\ref{ntot})).

Comparing (\ref{eq:final_result1}) with (\ref{eq:final_result}) we introduce the cross section averaged over the parton densities

\begin{equation}
\bar\sigma(E)=\frac{1}{2}\int \mathrm{d}x\left\langle \sigma(xs,Q^2)\right\rangle q(x,Q^2), \label{eq:f}
\end{equation}

which allows to rewrite the damping rate in the ordinary form

\begin{equation}
\gamma(E)=\bar \sigma(E)n_0. \label{eq:ff}
\end{equation}

For the reaction (\ref{s_channel}) one obtains  

\begin{equation}
\bar\sigma_{\nu e}(E)=\frac{1}{2}\int_0^\infty \mathrm{d}x\left\langle \sigma_{\nu e}(xs)f_{e}(x,s)\right\rangle, \label{eq:fi}
\end{equation}

where $\sigma_{\nu e}(s)$ is the cross section of the resonant subprocess $\nu_e e^+\rightarrow W^+$.
The integration over $\theta$ includes the lepton distribution function  since the latter, in this case, also depends on $s$ (we deal with a Drell--Yan-like process \cite{cmb_parton}).

\begin{figure}
\centering
\resizebox{0.5\textwidth}{!}{%
\includegraphics{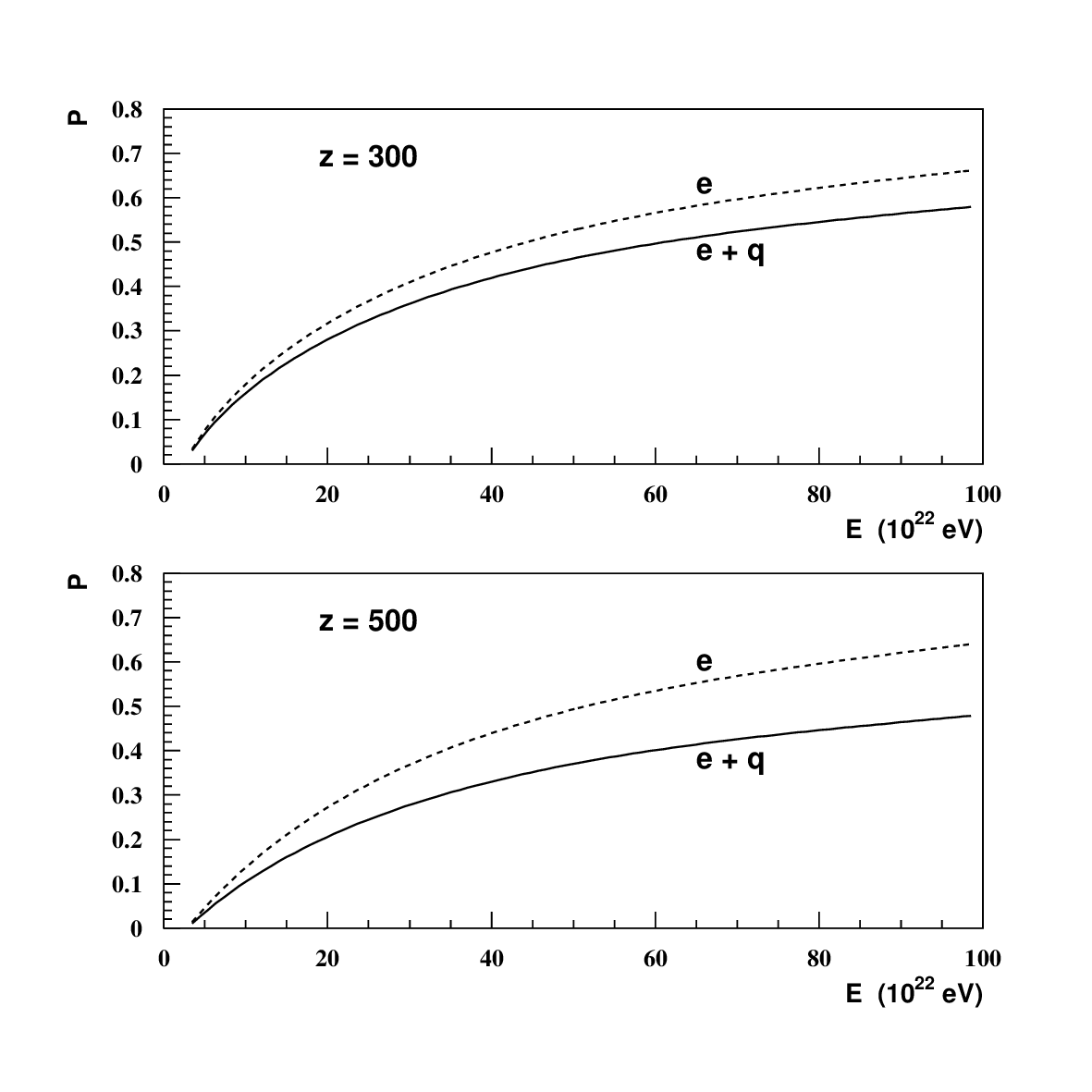}
}\caption{Dependence of the survival probability of an UHE electron neutrino emitted at $z_s=300$ (upper panel) and $500$ (lower panel) on its present day  energy $E$. The dashed curve is obtained under the assumption that the only contributing channel is $\nu_e\gamma\rightarrow W^+X$, while the solid one includes also the contribution of $\nu_e\gamma\rightarrow e^-X$}
\label{fig4}
\end{figure}

Knowing $\gamma(E)$, we can find the survival probability of an UHE neutrino traveling through the CMB radiation. For a constant damping, the probability as a function of $E$ and the propagation time $\tau$ is \cite{neutr_damping3} 

\begin{equation}
P(E,\tau)=e^{-\gamma(E)\tau}.\label{eq:prob_def}
\end{equation}

In fact, the travel without interaction can be so long that many cosmological quantities will have enough time to  significantly change due to the expansion of the universe (for example temperature of the CMB and therefore $n_0$). The neutrino energy will also get shifted to smaller values because of the expansion, just as in the case of light. This means that for a realistic situation the damping rate depends on time.  In order to take this dependence into account it is convenient to use the redshift $z$ which relates the present day values of the neutrino energy $E$ and the CMB temperature $T$ to those in the past. As we look back in time, these quantities increase with redshift as $E(1+z)$, $T(1+z)$. We can then generalize (\ref{eq:prob_def}) by integrating the damping rate over all redshifts from now ($z=0$) back to the neutrino source position $z_s$ \cite{neutr_damping2,neutr_damping3,neutr_damping4}:

\begin{equation}
P(E,z_s)=\exp\left[-\int_0^{z_s}\frac{\mathrm{d}z}{H(z)(1+z)}\gamma(E(1+z))\right],
\label{eq:prob_gen}
\end{equation}

where $\mathrm{d}\tau=-\mathrm{d}z/[H(z)(1+z)]$ with $H(z)$ being the Hubble parameter (see e.g. \cite{cosmology}). Note that the CMB temperature entering into (\ref{eq:prob_gen}) (though not explicitly indicated) is understood also to scale as $T(1+z)$. We take $H(z)=H_0\sqrt{0.3(1+z)^3+0.7}$ \cite{neutr_damping3} with $H_0=0.787\times10^{-28}$ cm$^{-1}$ \cite{pdg}.

We first calculate the survival probability of the UHE neutrinos emitted at $z_s=10, 15$ and $20$ regarding the reaction (\ref{s_channel}) as the leading absorption channel. We parameterize $\sigma_{\nu e}(s)$ in (\ref{eq:fi}) by the Breit--Wigner formula 

\begin{equation}
\sigma_{\nu e}(s)=24\pi\frac{\mathrm{\Gamma_{\nu e}}\mathrm{\Gamma}}{(s-m_W^2)^2+m_W^2\mathrm{\Gamma}^2},
\label{eq:2}
\end{equation}

where $m_W$ is the mass of the $W^+$ boson, ${\mathrm{\Gamma}}_{\nu e}$ is the partial
width for the decay $W^+\to\nu_e e^+$, and $\mathrm{\Gamma}$  is the total decay width of $W^+$. The corresponding results for the CMB temperature $T=2.725$ K are shown in Fig. \ref{fig3} (we set  $m_W=80.398$ GeV, ${\mathrm{\Gamma}}_{\nu e}=0.230$ GeV, $\mathrm{\Gamma}=2.141$ GeV and $\alpha(m_W^2)=1/128$ \cite{pdg} in the positron distribution function). One can see that the probability falls with increasing $z_s$ (as intuitively expected) and the dip due to the resonance absorption becomes more distinct. 

We have also found the contribution of the reaction (\ref{t_channel}) to the overall neutrino absorption. This turned out to be considerable at higher values of $z_s$.

It has been stated above that $Q^2$ in (\ref{eq:f}) is fixed. Meanwhile we have to take all possible four-momentum transfers into account. To do this we have  used the cross sections of the subprocesses of interest ($\nu_e d\to e^-u$ and $\nu_e\bar u\to e^-\bar d$) in the form

\begin{equation}
\frac{\mathrm{d}\sigma_{\nu d}}{\mathrm{d}Q^2}=\frac{G_F^2}{\pi}\frac{m_W^4}{(m_W^2+Q^2)^2}, \label{eq:fin}
\end{equation}

\begin{equation}
\frac{\mathrm{d}\sigma_{\nu \bar u}}{\mathrm{d}Q^2}=\frac{G_F^2}{\pi}\frac{m_W^4}{(m_W^2+Q^2)^2}\frac{(s-Q^2)^2}{s^2}, \label{eq:fina}
\end{equation}
where $G_F$ is the Fermi coupling constant.

After replacing $\sigma(x,Q^2)$ in (\ref{eq:f}) by $\mathrm{d}\sigma_{\nu q}/\mathrm{d}Q^2$ and substituting thus obtained damping rates into (\ref{eq:prob_gen}) we have integrated over $Q^2$ (from 10 GeV$^2$ to $xs$) besides the other integrations. The lower integration limit is chosen to be 10 GeV$^2$ since we are interested in neutrino--photon interactions which can produce final state hadrons. 
The corresponding results for $T=2.725$ K, $z_s=300$ and 500 are shown in Fig. \ref{fig4} (we set  $G_F=1.16637\times10^{-5}$ GeV$^{-2}$, and $\alpha=1/137.035$ \cite{pdg} in the quark distribution functions). One can see that the contribution of the channel (\ref{t_channel}) grows with the neutrino energy $E$ as well as with the distance from the source $z_s$. It is also notable that almost all the neutrinos emitted at $z_s=500$ with energies about $E(1+z_s)=25\times10^{24}$ eV would be absorbed by the CMB.

\section{Conclusions}{\label{conclusions}}
We have discussed how parton distributions can be attributed to a photon gas. We have regarded the CMB radiation as a particular case and noted that its parton content could be resolved by UHE cosmic neutrinos. We have considered the possibility of absorption of the UHE neutrinos by the CMB via the channels (\ref{s_channel}) and (\ref{t_channel}).  According to our calculations the universe turns out to be quite opaque to the neutrinos due to the presence of the CMB radiation. For example,  more than 10 \% of the UHE neutrinos emitted at $z_s=15$ with the initial energy $10^{25}$~eV would be absorbed. Moreover, the absorption process is expected to be accompanied by hadron production thus providing a mechanism of generation of UHE cosmic rays. It is interesting to compare our results with the predictions of the survaval probability of the UHE neutrinos traversing the relic neutrino background (C$\nu$B) (see e.g. \cite{neutr_damping1,neutr_damping2,neutr_damping3,neutr_damping4}).  It is essential that the C$\nu$B has never been detected directly and there are only theoretical estimations of its parameters (such as temperature) while the properties of the CMB have been well established experimentally. Therefore in studying the C$\nu$B by observing the resonant annihilation of the UHE neutrinos on the background antineutrinos through the reaction $\nu\bar\nu\to Z^0$ one has to be able to distinguish between signals from  this reaction and the $\nu\gamma$ interactions. This is possible, at least in principle. 
We have taken into account only the contribution of the charged current deep inelastic neutrino--photon scattering since in such a process, apart from the hadrons in the final state, the charged lepton is produced (in our case this is the electron). The latter being observed in correlation with the hadrons (or their decay products) may serve as an additional signature of the UHE neutrinos.

\begin{acknowledgement}
I thank F. F. Karpeshin for highlighting a few important
points on which I have focused more attention. This work was
supported in part by the Russian Foundation for Basic Research
(grant 06-02-16135).
\end{acknowledgement}



\begin{thebibliography}{}
%

%
\bibitem{feynman}
R.P.~Feynman \textit{Photon--Hadron Interactions}, (Benjamin, New
York, 1972)
%

\bibitem{review_phot1}
R.~Nisius, Phys. Rep. {\bf 332}, 165 (2000)

\bibitem{review_phot2}
M.~Krawczyk, M.~Staszel, A.~Zembrzuski, Phys. Rep. {\bf 345}, 265
(2001)

\bibitem{ref:dis_photon1}
S.J.~Brodsky, T.~Kinoshita, H.~Terazawa, Phys. Rev. D {\bf 4},
1532 (1971)

\bibitem{ref:dis_photon2}
T.F.~Walsh, P.~Zerwas, Nucl. Phys. B {\bf 41}, 551 (1972)

\bibitem{ref:log_dep1}
T.F.~Walsh, P.~Zerwas, Phys. Lett. B {\bf 44}, 195 (1973)

\bibitem{ref:log_dep2}
R.L.~Kingsley, Nucl. Phys. B {\bf 60}, 45 (1973)


\bibitem{ref:qcd}
E. Witten, Nucl. Phys. B {\bf 120}, 189 (1977)


\bibitem{gluk3}
M.~Gl\"uck, E.~Reya, A.~Vogt, Phys.\ Rev.\ D {\bf 45}, 3986 (1992)

\bibitem{review_phot3}
P.~Aurenche, M.~Fontannaz, J.Ph.~Guillet, Eur. Phys. J. C {\bfseries 44}, 395 (2005)

\bibitem{review_phot4}
W.~Slominski, H.~Abramowicz, A.~Levy, Eur. Phys. J. C {\bfseries 45}, 633 (2006)

\bibitem{cmb_spectr} D.J.~Fixsen et al., Astrophys. J. {\bfseries 473}, 576 (1996)

\bibitem{greisen}
K.~Greisen,
Phys.\ Rev.\ Lett. {\bf 16}, 748 (1966)
%
\bibitem{zatsepin}
G.T.~Zatsepin, V.A.~Kuzmin, Sov. Phys.\  JETP \ Lett.\ {\bf 4}, 78
(1966)

\bibitem{sigl}
P. Bhattacharjee, G. Sigl,
Phys.\ Rep.\ {\bf 327}, 109 (2000)

\bibitem{cmb_parton}
I.~Alikhanov, Eur. Phys. J. C {\bfseries 56}, 479 (2008); Erratum ibid. C {\bfseries 60}, 691 (2009)


\bibitem{greiner}
W.~Greiner, S.~Schramm, E.~Stein, \textit{Quantum Chromodynamics},
2nd edn. (Springer, Berlin, 2002)
%


\bibitem{landau}
L.D.~Landau, E.M.~Lifshitz, \textit{Statistical Physics}, Part 1,
3rd edn. (Pergamon Press, Oxford, 1980)
%

\bibitem{zeta}
E.C.~Titchmarsh, D.R.~Heath-Brown, \textit{The Theory of the Riemann
Zeta-Function}, 2nd edn. (Clarendon Press, Oxford, 1986)

\bibitem{seckel}
D. Seckel, Phys. Rev. Lett. {\bf 80},
900 (1998)

\bibitem{neutr_damping1}
P. Gondolo, G. Gelmini, S. Sarkar,  Nucl. Phys. B {\bfseries 392}, 111 (1993)

\bibitem{neutr_damping2}
G. Barenboim, O. Mena Requejo, C. Quigg, Phys. Rev. D {\bfseries 71}, 083002 (2005)

\bibitem{neutr_damping3}
J.C. D'Olivo, L. Nellen, S. Sahu, V. Van Elewyck, Astropart. Phys. {\bfseries 25}, 47  (2006)



\bibitem{neutr_damping4}
A. Ringwald, L. Schrempp, JCAP {\bfseries 0610}, 012 (2006)

\bibitem{close}
F.E. Close, \textit{An Introduction to Quarks and Partons}, (Academic Press, London, 1979)


\bibitem{cosmology}
S. Weinberg, \textit{Cosmology}, (Oxford University Press, New York, 2008)

\bibitem{pdg}
C. Amsler et al., Phys. Lett. B {\bf 667}, 1 (2008)


\end{thebibliography}
\end{document}